\documentclass{elsart}
\begin{document}
\mbox{} \hfill UCLA/96/TEP/16
\begin{frontmatter}
\title{Self-dual Yang--Mills Theory and One-Loop Maximally Helicity
    Violating Multi-Gluon Amplitudes\thanksref{support}}
\thanks[support]{Partially supported by the NSF under contract
    PHY-92-18990.}

\author{Daniel Cangemi\thanksref{email}}
\address{ Department of Physics, University of California Los
    Angeles, Box 951547, Los Angeles, CA 90095-1547 }
\thanks[email]{cangemi@physics.ucla.edu}

\begin{abstract}%
  A scalar cubic action that classically reproduces the self-dual
  Yang--Mills equations is shown to generate one-loop QCD amplitudes
  for external gluon all with the same helicity. This result is
  related to the symmetries of the self-dual Yang--Mills equations.
\end{abstract}

\begin{keyword}
    Integrable, Yang--Mills, helicity, amplitude, self-dual
\end{keyword}
\end{frontmatter}

\setlength{\unitlength}{1mm}

The study of multi-jet processes requires the calculation of QCD
amplitudes with external quarks and gluons. The complexity of the
calculations beyond tree level has motivated the development of
ingenious and efficient methods that involve spinor helicity, color
decomposition, supersymmetry, string theory, recursion relations,
factorization and unitarity~\cite{bern}.  The easiest amplitudes to
consider are the ones with all external gluons (conventionally taken
as outgoing) in the same polarization
state~\cite{parke,berends,mangano}. These so-called maximally helicity
violating (MHV) processes vanish at tree level and take at one-loop a
non-zero but very simple form. Zero tree amplitudes are a typical
signature of a quantized integrable system~\cite{ooguri,korepin} and
it is tempting to look for an integrable model that reproduces these
results. The relevance of two-dimensional current algebra for some MHV
amplitudes was already noticed by Nair some time ago~\cite{nair}. In a
recent talk~\cite{bardeen}, Bardeen showed that some solutions of the
self-dual Yang--Mills (SDYM) equations naturally appear in the
calculation of tree MHV amplitudes (see also \cite{selivanov}).
He also conjectured that the vanishing of the tree amplitudes is a
consequence of the symmetries of the SDYM system and that an anomaly
is responsible for the structure of the one-loop amplitudes.

In this paper, I extend Bardeen's analysis and obtain the one-loop MHV
QCD amplitudes from a scalar action whose Euler--Lagrange equations
reproduce the SDYM ones. In the first section, I briefly review the
calculation of QCD amplitudes. In section~2, I consider the SDYM
equations and two actions associated to them. In section~3, I
follow Bardeen's idea and show the relation between SDYM solutions and
MHV amplitudes. I extend this analysis to one-loop in section~4 and
discuss in section~5 the symmetries of the SDYM system. I conclude in
the last section with some remarks.

\section{A brief review of QCD amplitudes}

This section reviews some of the basic ingredients which simplify
the calculation of QCD amplitudes. I only present the material
relevant to my discussion and I refer the reader to Ref.~\cite{bern}
and references therein for a more complete presentation.

The spinor helicity notation~\cite{spinor} compactifies otherwise
lengthy expressions. An on-shell momentum $k$ (this always means
$k^2=0$) has three independent components that are usually expressed
in terms of scalar products of massless Weyl spinors (see Appendix~A).
It is more natural in this paper to choose\footnote{I use the
  conventions $k_{0\pm z} = k_0 \pm k_z$ and $k_{x\pm \mathrm{i}y} =
  k_x \pm \mathrm{i} k_y$, with metric $g_{\mu\nu} =
  \mathrm{diag}(1,-1,-1,-1)$ and totally antisymmetric tensor
  $\epsilon_{0123} = 1$.} $k_{0-z}$, $k_{x+\mathrm{i}y}$ and the ratio
\begin{equation}
  Q = \frac{ k_{0+z} }{ k_{x+\mathrm{i}y} } = \frac{ k_{x-\mathrm{i}y}
    }{ k_{0-z} }.
\end{equation}
The combination
\begin{equation}
  \label{Xdef}
  X(k_1,k_2) = k_{1,x+\mathrm{i}y} k_{2,0-z} - k_{1,0-z}
  k_{2,x+\mathrm{i}y} = (Q_1 - Q_2)^* \; k_{1,0-z} k_{2,0-z}
\end{equation}
satisfies the identity
\begin{equation}
  \label{Xprop}
  2\; k_1 \cdot k_2 = X(1,2) \; (Q_1 - Q_2) \,.
\end{equation}
One also introduces a set of polarization vectors in an arbitrary
reference frame labelled by the null vector $q$ ($q \cdot k \not = 0$).
A change in $q$ is analogous to a gauge transformation of the external
legs and since the amplitudes are gauge invariant, they do not depend on the
reference frame chosen for the external particles. Nevertheless the
calculations may drastically simplify for some reference vectors $q$.
These polarization vectors are orthogonal to the momentum $k$ and
normalized in the following way:
\begin{equation}
  k \cdot \varepsilon^\pm(k;q) = 0 \,, \quad \varepsilon^+(k;q) \cdot
  \varepsilon^-(k;q) = -1 \,, \quad \varepsilon^\pm(k_1;q) \cdot
  \varepsilon^\pm(k_2;q) = 0 \,.
\end{equation}
Their general form is given in Appendix~A.
Comparison with the SDYM system is made later on in the ``light-cone
gauge'' $q_\mu = (1, 0, 0, 1)$, in which the polarization vectors are
denoted $\varepsilon^{(\pm)}_\mu(k)$ and have the non zero components
\begin{eqnarray}
  \varepsilon^{(+)}_{0+z}(k) = - \sqrt{2} \frac{k_{0+z}}{k_{x+\mathrm{i}y}}
  \,,
  &&\qquad \varepsilon^{(+)}_{x-\mathrm{i}y}(k) = - \sqrt{2}
     \frac{k_{x-\mathrm{i}y}}{k_{x+\mathrm{i}y}} \,, \nonumber\\[-6pt]
  \label{polarization} \\[-6pt]
  \varepsilon^{(-)}_{0+z}(k) = - \sqrt{2}
  \frac{k_{0+z}}{k_{x-\mathrm{i}y}} \,,
  &&\qquad \varepsilon^{(-)}_{x+\mathrm{i}y}(k) = - \sqrt{2}
     \frac{k_{x+\mathrm{i}y}}{k_{x-\mathrm{i}y}} \,. \nonumber
\end{eqnarray}
A positive helicity gauge field is parallel to $\varepsilon^{(+)}_\mu$:
\begin{eqnarray}
  &\displaystyle \varepsilon^{(+)}(k) \cdot A(k) = - \frac{1}{\sqrt{2}}
  \frac{1}{k_{x+\mathrm{i}y}} \Bigl( k_{0+z} A_{0-z}(k) -
  k_{x-\mathrm{i}y} A_{x+\mathrm{i}y}(k) \Bigr) = 0 \,,
  \nonumber\\[-6pt]
  \label{helicity} \\[-6pt]
  &2 \; k \cdot A(k) = k_{0-z} A_{0+z} + k_{0+z} A_{0-z} -
  k_{x+\mathrm{i}y} A_{x-\mathrm{i}y} - k_{x-\mathrm{i}y}
  A_{x+\mathrm{i}y} = 0 \,. \nonumber
\end{eqnarray}
We will see in the next section that a self-dual gauge field in the
light-cone gauge has positive helicity.

Simple $N=1$ and $N=2$ supersymmetry arguments~\cite{tasi,andrew}
relate one-loop MHV amplitudes with either gluons, scalars or fermions
running in the loop. For $\mathrm{SU}(N_c)$ gauge fields, $n_s$ real
scalar fields in the representation $R_s$ and $n_f$ Dirac fermions
(four components) in the representation $R_f$, the amplitude has an
overall factor of\footnote{I use the following group theoretic
  conventions for $\mathrm{SU}(N_c)$: $C_2(G) = 2 N_c \Leftrightarrow
  [ T^a, T^b ] = \mathrm{i} \sqrt{2} f^{abc} T^c$, $T_2(F) = 1
  \Leftrightarrow \;\mathrm{tr}_F\; T^a T^b = \delta^{ab}$.}
\begin{equation}
  \label{common}
  \left( C_2(G) + \frac{n_s}{2} \; T_2(R_s) - 2 \, n_f \; T_2(R_f)
  \right) \,.
\end{equation}
Thus, it is sufficient to consider an (adjoint) scalar field $\chi$
coupled to a Yang--Mills field,
\begin{equation}
  \label{action-QCD}
  \mathcal{L}_{\mathrm{QCD}} = - \frac{1}{4} \;\mathrm{tr}\; F_{\mu\nu}
  F^{\mu\nu} + \frac{1}{2} \;\mathrm{tr}\; D_\mu \chi D^\mu \chi \,,
\end{equation}
where the covariant derivative is $D_\mu = \partial_\mu - \mathrm{i}g
A_\mu / \sqrt{2}$ and the field strength is $F_{\mu\nu} = \partial_\mu
A_\nu - \partial_\nu A_\mu - \mathrm{i} g [ A_\mu, A_\nu ] /
\sqrt{2}$, and to calculate one-loop amplitudes with classical gauge fields
and scalar fields only in the loop.
%
%
The scalar propagator is $\Delta = \mathrm{i}/k^2$ with a color factor
$\delta^{ab} / C_2(G)$.

The color structure of QCD amplitudes may be simplified by decomposing
them in color-ordered partial amplitudes~\cite{color}: At one-loop, an
$(n+1)$ gluon amplitude with external momenta $k_1, \ldots, k_{n+1}$,
helicities $\varepsilon_1, \ldots, \varepsilon_{n+1}$ and color
indices $a_1, \ldots, a_{n+1}$ is written as
\begin{eqnarray}
  \lefteqn{
  M^\mathrm{one-loop}_{n+1}(\{k_i, \varepsilon_i, a_i\}) =} \hspace{1cm}
   \nonumber\\
   & \sum_{{\mathrm{non\ cyclic} \atop \mathrm{permutations}} \atop
     \mathrm{of\ } (1\cdots n+1)} \;\mathrm{tr}\;
   ( T^{a_1} \cdots T^{a_{n+1}} ) \;
   m^\mathrm{one-loop}(k_1,\varepsilon_1 ; \ldots ;
     k_{n+1},\varepsilon_{n+1}) \nonumber + \ldots \,.
\end{eqnarray}
The triple dots at the end represent additional sub-leading color
structures that can be calculated from the above leading-color partial
amplitudes~\cite{dixon}. In other words, it is sufficient to calculate
the fewer partial amplitudes using color-ordered Feynman rules where
color indices are absent.

Amplitudes with all, or all but one, external gluons with the same
helicity identically vanish at tree level:
\begin{eqnarray}
  \label{mtree}
  \lefteqn{
  m^{\mathrm{tree}}(k_1,+; \cdots, k_n,+; k_{n+1},\pm) = \mathrm{i}
  k_1^2 \; \varepsilon^{+,\mu_1}(k_1;q_1) \; \cdots \; \mathrm{i}
  k_n^2 \; \varepsilon^{+,\mu_n}(k_n;q_n) \; }
  \nonumber \\
  &&{} \times \mathrm{i} k_{n+1}^2 \;
    \varepsilon^{\pm,\mu_{n+1}}(k_{n+1};q_{n+1}) \left.
    \; \Bigl\langle A_{\mu_1}(k_1) \cdots A_{\mu_{n+1}}(k_{n+1})
    \Bigr\rangle^\mathrm{tree}_c \; \right|_{k_1^2 = \cdots = k_{n+1}^2 = 0}
    = 0 \,. \nonumber\\
\end{eqnarray}
The subscript $c$ stands for connected Green functions. This can be
proven using the freedom one has in the choice of the reference
momenta $q_i$ for the external particle helicities~\cite{tasi}.
Alternatively one can show that the current amplitude,
\begin{eqnarray}
  \label{QCD-current}
  \lefteqn{
  \Bigl\langle A_\mu(k) \Bigr\rangle_{1^+ \ldots n^+}
  \equiv \varepsilon^{(+),\mu_1}(k_1) \cdots \varepsilon^{(+),\mu_n}(k_n)
  } \qquad \nonumber\\
  &&{}\times (\mathrm{i} k_1^2) \cdots (\mathrm{i}
  k_n^2) \left. \Bigl\langle A_{\mu_1}(k_1) \cdots A_{\mu_n}(k_n)
    A_\mu(k) \Bigr\rangle_c \;  \right|_{k_1^2 = \cdots = k_n^2 = 0} \,,
\end{eqnarray}
has no multi-particle poles at tree level. We will later calculate the
current amplitudes (\ref{QCD-current}) using a self-dual system
and demonstrate that they have no pole in $k^2$.

At one-loop, MHV amplitudes with four and more external legs do not vanish
and are explicitly known.  For example, the four particle
amplitude is~\cite{david}
\begin{eqnarray}
  \label{QCD-1L}
  \lefteqn{
  m^{\mathrm{one-loop}}(k_1,+; \cdots k_4,+) =} \qquad \nonumber\\
  &&= - N_c \, g^4 \frac{\mathrm{i}}{48 \pi^2} \;
  \frac{k_{1,x+\mathrm{i}y}}{Q_1} \frac{k_{2,x+\mathrm{i}y}}{Q_2}
  \frac{k_{3,x+\mathrm{i}y}}{Q_3} \frac{k_{4,x+\mathrm{i}y}}{Q_4} \;
  \frac{X(k_1+k_2,k_3)^2}{(Q_1 - Q_2)^2} \,,
\end{eqnarray}
A more conventional expression written in spinor notations is given
in Appendix~A. It is the goal of this paper to reproduce all the
MHV one-loop amplitudes in the framework of a SDYM system.

\section{SDYM equations}

The SDYM equations are most commonly studied in Euclidean space or
$(2+2)$-dimensional spacetime so that the solutions are real. In
Euclidean space, SDYM solutions describe instantons and in a $(2+2)$
signature, they describe the gauge sector of an $N=2$ heterotic or
open string~\cite{ooguri}. Here I explicitly work in Minkowski
spacetime and consider the self-dual equations,
\begin{equation}
  \label{SDYM}
  F_{\mu\nu} = \frac{\mathrm{i}}{2} \epsilon_{\mu\nu\rho\sigma}
  F^{\rho\sigma} \,,
\end{equation}
whose solutions necessarily live in the complexification of the gauge
group. Complex self-dual configurations have nevertheless a physical
interpretation as waves of positive helicity. For Maxwell-like fields
($g = 0$, $E^j = F_{0j}, B^j = \epsilon^{0jkl} F_{kl} / 2$), these
equations simply state that electric and magnetic fields form a
positive helicity wave, $E^j = \mathrm{i} B^j$.  In components,
Eqns.~(\ref{SDYM}) read
\begin{equation}
  \label{sdym}
  F_{0+z,x-\mathrm{i}y} = 0 \,, \qquad
  F_{0-z,x+\mathrm{i}y} = 0 \,, \qquad
  F_{0-z,0+z} = F_{x+\mathrm{i}y,x-\mathrm{i}y} \,.
\end{equation}
Considerable effort has been devoted to the study of these equations
and various approaches have been proposed.

One approach comes from the original work of Yang~\cite{yang} and
views the two first equations as zero curvature conditions for the
fields $(A_{0+z}, A_{x-\mathrm{i}y})$ and $(A_{0-z},
A_{x+\mathrm{i}y})$.  They are solved using two elements $h, \bar h$
of the (complexified) gauge group:
\begin{eqnarray}
  &-\frac{\mathrm{i}g}{\sqrt{2}} A_{0+z} = h^{-1} \partial_{0+z} h \,, \qquad
   -\frac{\mathrm{i}g}{\sqrt{2}} A_{x-\mathrm{i}y} =
     h^{-1} \partial_{x-\mathrm{i}y} h \,,
  \nonumber \\[-6pt]
  \\[-6pt]
  &-\frac{\mathrm{i}g}{\sqrt{2}} A_{0-z} =
     \bar h^{-1} \partial_{0-z} \bar h \,, \qquad
  -\frac{\mathrm{i}g}{\sqrt{2}} A_{x+\mathrm{i}y} =
     \bar h^{-1} \partial_{x+\mathrm{i}y} \bar h \,. \nonumber
\end{eqnarray}
The last self-dual equation involves the product $H = h
\bar h^{-1}$,
\begin{equation}
  \partial_{0-z} ( H^{-1} \partial_{0+z} H ) -
  \partial_{x+\mathrm{i}y} ( H^{-1} \partial_{x-\mathrm{i}y} H ) = 0 \,,
\end{equation}
and is known as the Yang equation. This is the Euler--Lagrange
equation of an action proposed by Donaldson~\cite{donaldson} and by
Nair and Schiff~\cite{schiff}:
\begin{eqnarray}
  \label{actionJ}
  S_{\mathrm{DNS}}(H) &=& \frac{f_\pi^2}{2} \int \d^4x \;\mathrm{tr}\;
  \left( \partial_{0+z} H \partial_{0-z} H^{-1} -
    \partial_{x-\mathrm{i}y} H \partial_{x+\mathrm{i}y} H^{-1} \right)
  \nonumber\\
  &&{} + \frac{f_\pi^2}{2} \int \d^4x \d t
    \;\mathrm{tr}\; \left( \, [ H^{-1} \partial_{0+z} H, H^{-1}
    \partial_{0-z} H ] - \right.\nonumber\\
  &&{} \hspace{3.5cm} \left. [ H^{-1} \partial_{x-\mathrm{i}y} H, H^{-1}
    \partial_{x+\mathrm{i}y} H ] \; \right)  H^{-1} \partial_t H \,.
\end{eqnarray}
The constant $f_\pi$ has mass dimension one. This action is expected
to be the prototype of a four-dimensional conformal theory that
generalizes the two-dimensional Wess--Zumino--Novikov--Witten
model~\cite{moore}.  Namely, a direct calculation shows that its beta
function vanishes at one-loop order. ( It is even argued that it
vanishes at all orders in some circumstances~\cite{ketov}. )

We will follow another approach~\cite{bruschi}, that takes advantage
of the light-cone gauge $A_{0-z} = 0$. In this gauge, the two last
equations in~(\ref{sdym}) are equivalent to the positive helicity
conditions~(\ref{helicity}). The second equation implies
$A_{x+\mathrm{i}y} = 0$ and the third one is an integrability
condition for the fields $A_{0+z}, A_{x-\mathrm{i}y}$,
\begin{equation}
  \label{A-Phi-a}
  A_{x+\mathrm{i}y} = 0 \,, \qquad A_{0+z} = \sqrt{2} \;
  \partial_{x+\mathrm{i}y} \Phi \,, \qquad A_{x-\mathrm{i}y} =
  \sqrt{2} \; \partial_{0-z} \Phi \,,
\end{equation}
or, in momentum space,
\begin{equation}
  \label{A-Phi}
  \begin{array}{l}
    A_{0+z}(k) = - \mathrm{i} \sqrt{2} \, k_{x+\mathrm{i}y} \; \Phi(k) \\
    A_{x-\mathrm{i}y}(k) = - \mathrm{i} \sqrt{2} \, k_{0-z} \; \Phi(k)
  \end{array}
  \quad \rlap{\ \ $\Longrightarrow$} \raisebox{.8em}{$k^2 = 0$}
  \quad A_\mu(k) = \varepsilon^{(+)}_\mu(k) \; \mathrm{i}
  \frac{k_{x+\mathrm{i}y}}{Q} \; \Phi(k) \,.
\end{equation}
A self-dual gauge field has positive helicity. The scalar field $\Phi$ is
constrained by the remaining first equation in~(\ref{sdym}):
\begin{equation}
  \label{equPhi}
  \partial^2 \Phi - \mathrm{i} g \; [ \partial_{x+\mathrm{i}y} \Phi,
  \partial_{0-z} \Phi ] = 0 \,.
\end{equation}
These are the equations of motion of the scalar action~\cite{parkes}:
\begin{equation}
  \label{actionPhi}
  S_{\mathrm{scalar}}(\phi) = f_\pi^2 \int \d^4x \;\mathrm{tr}\;
  \left( \frac{1}{2} \partial \phi \cdot \partial \phi +
    \frac{\mathrm{i} g}{3} \; \phi \; [ \partial_{x+\mathrm{i}y} \phi,
    \partial_{0-z} \phi ] \right) \,.
\end{equation}
As before, $f_\pi$ has mass dimension one so that the complex field
$\Phi$ is dimensionless, in agreement with~(\ref{A-Phi-a}). This
action has three unconventional properties: it is not
real,\footnote{One can of course add its complex conjugate. The
  kinetic term does not have the standard form and the fields $\Phi$
  and $\Phi^*$ are totally decoupled.} it explicitly breaks Lorentz
invariance and it is not renormalizable by power counting since the
interaction term contains two derivatives.  Nevertheless, it turns out
to reproduce correctly the MHV amplitudes up to one loop.

\section{Tree amplitudes}

I first recall the relation between a classical solution and a tree
amplitude. An amplitude is obtained by truncating on-shell a connected
Green function. A tree Green function is generated by the Legendre
transform $W(J)$ of the classical action $S(\phi)$:
\begin{eqnarray}
  & \frac{\delta S(\phi)}{\delta \phi(x)} \biggr|_{\phi=\Phi_J}
      + J(x) = 0 \,, \\
  & W(J) = S( \Phi_J ) + \int dx \; \Phi_J(x) J(x) \,,\\
  & \langle \phi(x_1) \cdots \phi(x_{n+1}) \rangle_c = \frac{ \mathrm{i}
      \delta^{n+1} W(J) }{ \mathrm{i} \delta J(x_1) \cdots \mathrm{i}
      \delta J(x_{n+1}) } \; \biggr|_{J=0} \,.
\end{eqnarray}
Since the classical solution $\Phi_J$ in presence of a source $J$ is given by
the first variation of the generating functional $W(J)$, we have
\begin{equation}
  \label{funct}
  \langle \phi(x_1) \cdots \phi(x_{n+1}) \rangle_c = \frac{ \delta^{n}
  \Phi_J(n+1) }{ \mathrm{i} \delta J(x_1) \cdots\mathrm{i}  \delta
  J(x_n) } \; \biggr|_{J=0}  \,.
\end{equation}
The classical solution $\Phi_J$ is an infinite series in $J$ whose
coefficients are the connected Green functions.

In our case, the equations of motion with source are
\begin{equation}
  \label{equPhiJ}
  \partial^2 \Phi_J - \mathrm{i} g \; [ \partial_{x+\mathrm{i}y} \Phi_J,
  \partial_{0-z} \Phi_J ] + J = 0 \,.
\end{equation}
These equations are solved with a Bethe \textit{Ansatz}~\cite{bardeen}.
The solution is iteratively obtained as a series in the coupling
constant $g$:
\begin{equation}
  \Phi_J(x) = \sum_{m=1}^\infty \Phi_J^{(m)}(x) \,, \qquad
  \Phi_J^{(m)} \propto g^{m-1} \,.
\end{equation}
In momentum space, the first term is
\begin{equation}
  \Phi_J^{(1)}(k) = J(k) / k^2 \equiv j(k) \,.
\end{equation}
It is possible to add to $\Phi_J^{(1)}(k)$ a function with support on
the light-cone $k^2=0$. This would shift the solution with source by a
solution of Eq.~(\ref{equPhi}) and lead to Green functions in the
presence of a non-vanishing background field.  Solutions of the SDYM
equations~(\ref{equPhi}) without source are obtained from $\Phi_J$ by
taking the support of $j(k) = J(k) / k^2$ on the light-cone. By
iteration we get
\begin{eqnarray}
  \label{recrel}
  \Phi_J^{(m)}(k) &=& i g \sum_{j=1}^{m-1} \int \frac{d^4p_1}{(2\pi)^4}
    \; \frac{d^4p_2}{(2\pi)^4} (2\pi)^4\delta(p_1 + p_2 - k)
  \nonumber\\
  && \hspace{3cm} {} \times \frac{X(p_1, p_2)}{(p_1 + p_2)^2} \;\;
     \Phi^{(j)}(p_1) \; \Phi^{(m-j)}(p_2) \,.
\end{eqnarray}
Each term $\Phi_J^{(m)}(k)$ is a sum of trees with cubic vertices,
attached to a leg with momentum $k$ and ending up on $m$ sources $J(p)
/ p^2 = j(p)$.
Since the current amplitude
\begin{eqnarray}
  \label{currampl}
  \Bigl\langle \phi(k) \Bigr\rangle_{1 \cdots n} \equiv
  (- \mathrm{i} k_1^2) f(k_1) \cdots (- \mathrm{i} k_n^2) f(k_n)
  \left. \Bigl\langle \phi(k_1) \cdots \phi(k_n) \phi(k) \Bigr\rangle
  \right|_{k_1^2 = \cdots = k_n^2 = 0} \nonumber\\
\end{eqnarray}
( $f(k_j)$ is a function with support on the light-cone $k_j^2 = 0$ )
is obtained by differentiating the classical solution $n$ times with
respect to the source $J$ and by truncating on-shell $n$ external legs
( see (\ref{funct}) ), we can restrict the support of $j(k)$ to be on
the light-cone and then take the derivatives with respect to $j(k)$.
Then Eqns.~(\ref{recrel}) are equivalent to the Berends and Giele
recursion relations~\cite{berends,recursion,mahlon} and one shows by
induction:
\begin{eqnarray}
 \label{treePhiJ}
 \lefteqn{
  \Phi_J^{(m)}(k) = ( \mathrm{i} g )^{m-1} \int \frac{d^4p_1}{(2\pi)^4} \cdots
    \frac{d^4p_m}{(2\pi)^4} (2\pi)^4\delta(p_1 + \cdots + p_m - k)}
  \qquad \nonumber\\
  &\times j(p_1) \cdots j(p_m)  \;
    (Q_1 - Q_2)^{-1} (Q_2 - Q_3)^{-1} \cdots (Q_{m-1} - Q_m)^{-1} \,.
\end{eqnarray}

If one takes $j$ to be a sum of $n$ independent (on-shell) plane
waves,
\begin{equation}
\label{planew}
  j(x) = - \mathrm{i} \sum_{j=1}^n \; T^{a_j} \e^{-\mathrm{i} k_j x} f(k_j) \,,
\end{equation}
one gets the Bethe \textit{Ansatz} solution~\cite{bardeen} of the SDYM
equations. The relevant part in the calculation of the current
amplitude~(\ref{currampl}) involves $n$ $j$'s and so is contained in
$\Phi^{(n)}$:
\begin{eqnarray}
 \label{treePhi}
  \Phi^{(n)}(x) &=& - \mathrm{i} g^{n-1} \sum_{\mathrm{permutations}
    \atop \mathrm{of\ }( 1 \cdots n )} T^{a_1} \cdots
  T^{a_n} \; \e^{-\mathrm{i} (k_1 + \cdots + k_n)x} f(k_1) \cdots
  f(k_n) \nonumber\\
  &\times& (Q_1 - Q_2)^{-1} (Q_2 - Q_3)^{-1} \cdots (Q_{n-1} - Q_n)^{-1}
    \; + \cdots \,.
\end{eqnarray}
The piece given here explicitly involves all momenta $k_j$ ($j = 1,
\ldots, n$) and is equal to the current amplitude~(\ref{currampl}).
The additional terms denoted by the dots ( as well as the other
$\Phi^{(m)}, \; m\not=n$ ) correspond to tree diagrams with two or
more legs having the same momentum. They are described in Appendix~B.

The function $f(k)$ has not yet been specified. One remarks that for
$f(k) = Q / k_{x+\mathrm{i}y}$, $\Phi^{(1)}(k) = j(k)$
corresponds ( see ~(\ref{A-Phi}) ) to a positive helicity gauge field:
$A_\mu^{(1)}(k) = \sum_{j=1}^n (2\pi)^4 \delta(k_j - k)
\varepsilon^{(+)}_\mu(k_j)$. For this same function one finds a
remarkable identity~\cite{bardeen} between the tree SDYM current
amplitude~(\ref{currampl}) and the tree MHV current
amplitudes~(\ref{QCD-current}) in the light-cone gauge and in the
reference frame~(\ref{polarization}) as calculated for example
in~\cite{mahlon}:
\begin{eqnarray}
  \label{tree-id}
  \begin{array}{l}
    \Bigl\langle A_{0+z}(k) \Bigr\rangle_{1^+ \cdots n^+} \; \rlap{\ \
      $=$}\raisebox{1em}{tree} \; - \mathrm{i} \sqrt{2} \;
    k_{x+\mathrm{i}y} \; \Bigl\langle \phi(k) \Bigr\rangle_{1 \cdots n} \,, \\
    \Bigl\langle A_{x-\mathrm{i}y}(k) \Bigr\rangle_{1^+ \cdots n^+} \;
    \rlap{\ \ $=$}\raisebox{1em}{tree} \; - \mathrm{i} \sqrt{2} \;
    k_{0-z} \; \Bigl\langle \phi(k) \Bigr\rangle_{1 \cdots n} \,.
  \end{array}
\end{eqnarray}
( compare with the self-dual \textit{Ansatz}~(\ref{A-Phi}) ).  The
MHV amplitudes are obtained by contracting~(\ref{tree-id}) with
$\mathrm{i} k^2 \varepsilon^{\pm,\mu}(k;q)$ and taking the limit $k^2
\to 0$. A remarkable property of the solution~(\ref{treePhi}) is that
the multiparticle poles appearing in the branches of the tree
completely disappear in the final solution. In particular, there is no
pole in $k^2 = (k_1 + \cdots + k_n)^2$ and thus the tree
amplitudes~(\ref{mtree}) vanish.

An analysis similar to the one done in this section can be
carried~\cite{selivanov} with a third \textit{Ansatz} for the
self-dual gauge field, originally proposed by 't Hooft for instantons
solutions.

\section{One-loop amplitudes}

The color-ordered Feynman rules for the scalar
action~(\ref{actionPhi}) are very simple. In momentum space the
propagator is $\mathrm{i} / k^2$ and the cubic vertex is
\begin{equation}
  \label{vertex-phi}
  \raisebox{-8mm}{
  \begin{picture}(25,18)(-14,-9)
    \put(-5,0){\line(-1,0){5}}\put(0,0){\vector(-1,0){5}}
    \put(0,0){\vector(1,1){4}}\put(2.82,2.82){\line(1,1){4}}
    \put(0,0){\vector(1,-1){4}}\put(2.82,-2.82){\line(1,-1){4}}
    \put(-14,-1){$\phi$}\put(8,6){$\phi$}\put(8,-7.5){$\phi$}
    \put(1,-5){$\scriptstyle 1$}
    \put(.5,3){$\scriptstyle 2$}
    \put(-5,1){$\scriptstyle 3$}
  \end{picture}
  } = \frac{g}{2} \; X(k_3, k_1 - k_2) = g \; X(k_1,k_2) \,.
\end{equation}
Minkowski kinematics is such that this vertex is zero on-shell in
agreement with the vanishing of the three particle MHV
amplitudes. This on-shell vertex is not zero for
Euclidean and $(2+2)$ signature.  Although two and three particle
amplitudes at one-loop involve IR and UV divergent
integrals~\cite{parkes}, they vanish for similar kinematics reasons.
I have computed the color-ordered four particle amplitude using a
symbolic algebra computer program and performed the resulting
integrals in dimensional regularization~\cite{david}:
\begin{eqnarray}
  \lefteqn{
  \left. ( - \mathrm{i} k_1^2) \cdots ( - \mathrm{i} k_4^2) \;
    \Bigl\langle \phi(k_1) \cdots \phi(k_4) \Bigr\rangle
  \right|_{k_1^2 = \cdots = k_4^2 = 0} } \hspace{2in} \nonumber\\
  &&= - \frac{N_c \, g^4}{2} \frac{\mathrm{i}}{48 \pi^2} \frac{X(k_1 +
    k_2, k_3)^2}{(Q_1 - Q_2)^2} \,.
\end{eqnarray}
If one multiplies each leg by $- \mathrm{i} Q / k_{x+\mathrm{i}y}$,
like at tree level, one gets a full agreement with QCD, see
Eq.~(\ref{QCD-1L}) (remember the factor of two in~(\ref{common})
between complex scalars and gluons contributions ). In this section, I
show that this result generalizes to one-loop amplitudes with an
arbitrary number of external gluons.

As mentioned in the first section, one-loop QCD amplitudes may be
calculated from the scalar QCD action~(\ref{action-QCD}) with a
classical gauge field $A_\mu$ and a quantized scalar field $\chi$
running in the loop. In light-cone gauge, (\ref{action-QCD}) reads
\begin{eqnarray}
  \label{back-QCD}
  \lefteqn{
  S_{\mathrm{QCD}} = \int \d^4x \;\mathrm{tr}\; \Biggl( - \frac{1}{4}
  F_{\mu\nu} F^{\mu\nu} + \frac{1}{2} \partial \chi \cdot \partial
  \chi - \frac{\mathrm{i}g}{2\sqrt{2}} [ A_{0+z}, \chi ] \;
  \partial_{0-z} \chi } \nonumber\\
  &&+ \frac{\mathrm{i}g}{2\sqrt{2}} [
  A_{x-\mathrm{i}y}, \chi ] \; \partial_{x+\mathrm{i}y} \chi +
  \frac{\mathrm{i}g}{2\sqrt{2}} [ A_{x+\mathrm{i}y}, \chi ] \;
  \partial_{x-\mathrm{i}y} \chi + \frac{g^2}{4} [
  A_{x+\mathrm{i}y}, \chi ] [ A_{x-\mathrm{i}y}, \chi ] \Biggr)
  \,, \nonumber\\
\end{eqnarray}
and generates five color-ordered vertices coupling gauge fields and
scalar fields:
\begin{eqnarray}
  \label{vertex-QCD}
  &\raisebox{-8mm}{
  \begin{picture}(30,18)(-19,-9)
    \put(-5,0){\line(-1,0){5}}\put(0,0){\vector(-1,0){5}}
    \put(0,0){\vector(1,1){4}}\put(2.82,2.82){\line(1,1){4}}
    \put(0,0){\vector(1,-1){4}}\put(2.82,-2.82){\line(1,-1){4}}
    \put(-19,-1){$A_{0+z}$}\put(8,6){$\chi$}\put(8,-7.5){$\chi$}
    \put(1,-5){$\scriptstyle 1$}
    \put(.5,3){$\scriptstyle 2$}
    \put(-5,1){$\scriptstyle 3$}
  \end{picture}
  } = \frac{\mathrm{i}g}{2\sqrt{2}} \; ( k_1 - k_2 )_{0-z} \,, \nonumber\\
  &\raisebox{-8mm}{
  \begin{picture}(31,18)(-20,-9)
    \put(-5,0){\line(-1,0){5}}\put(0,0){\vector(-1,0){5}}
    \put(0,0){\vector(1,1){4}}\put(2.82,2.82){\line(1,1){4}}
    \put(0,0){\vector(1,-1){4}}\put(2.82,-2.82){\line(1,-1){4}}
    \put(-20,-1){$A_{x-\mathrm{i}y}$}
    \put(8,6){$\chi$}\put(8,-7.5){$\chi$}
    \put(1,-5){$\scriptstyle 1$}
    \put(.5,3){$\scriptstyle 2$}
    \put(-5,1){$\scriptstyle 3$}
  \end{picture}
  } = - \frac{\mathrm{i}g}{2\sqrt{2}} \; ( k_1 - k_2 )_{x+\mathrm{i}y}
  \,,  \nonumber\\[-6pt]
  \\[-6pt]
  &\displaystyle
  \raisebox{-8mm}{
  \begin{picture}(31,18)(-20,-9)
    \put(-5,0){\line(-1,0){5}}\put(0,0){\vector(-1,0){5}}
    \put(0,0){\vector(1,1){4}}\put(2.82,2.82){\line(1,1){4}}
    \put(0,0){\vector(1,-1){4}}\put(2.82,-2.82){\line(1,-1){4}}
    \put(-20,-1){$A_{x+\mathrm{i}y}$}\put(8,6){$\chi$}
    \put(8,-7.5){$\chi$}
    \put(1,-5){$\scriptstyle 1$}
    \put(.5,3){$\scriptstyle 2$}
    \put(-5,1){$\scriptstyle 3$}
  \end{picture}
  } = - \frac{\mathrm{i}g}{2\sqrt{2}} \; ( k_1 - k_2 )_{x-\mathrm{i}y}
  \,, \nonumber\\
  &\raisebox{-8mm}{
  \begin{picture}(29,18)(-18,-9)
    \put(0,0){\vector(1,1){4}}\put(2.82,2.82){\line(1,1){4}}         
    \put(0,0){\vector(1,-1){4}}\put(2.82,-2.82){\line(1,-1){4}}      
    \put(0,0){\vector(-1,1){4}}\put(-2.82,2.82){\line(-1,1){4}}      
    \put(0,0){\vector(-1,-1){4}}\put(-2.82,-2.82){\line(-1,-1){4}}   
    \put(8,6){$\chi$}\put(8,-7.5){$\chi$}
    \put(-18,6){$A_{x\pm \mathrm{i}y}$}
    \put(-18,-7.5){$A_{x\mp \mathrm{i}y}$}
    \put(4,-3){$\scriptstyle 1$}\put(4.5,2){$\scriptstyle 2$}
    \put(-6,2){$\scriptstyle 3$}\put(-6,-3){$\scriptstyle 4$}
  \end{picture}
  } = - \frac{\mathrm{i} g^2}{4} \,. \nonumber
\end{eqnarray}
Gluon vertices are not shown here since they have already been
correctly taken into account in the previous section. One observes
that this action looks very similar to the SDYM
action~(\ref{actionPhi}) in a background field $\Phi$,
\begin{eqnarray}
  \lefteqn{
  S_{\mathrm{scalar}}(\Phi + \phi) = S(\Phi) + \int \d^4x
  \;\mathrm{tr}\; \Biggl( \frac{1}{2} \partial \phi \cdot \partial
  \phi - \frac{\mathrm{i} g}{2} [ \partial_{x+\mathrm{i}y} \Phi, \phi
  ] \partial_{0-z} \phi } \qquad \nonumber\\
  &&+ \frac{\mathrm{i} g}{2} [ \partial_{0-z}
  \Phi, \phi ] \partial_{x+\mathrm{i}y} \phi
  - \phi \; ( \partial^2 \Phi + \mathrm{i} g [
  \partial_{x+\mathrm{i}y} \Phi, \partial_{0-z} \Phi] ) \nonumber\\
  &&+ \frac{\mathrm{i}g}{3} \phi \; [ \partial_{x+\mathrm{i}y} \phi,
  \partial_{0-z} \phi] \Biggr) \,,
\end{eqnarray}
when one uses the relation~(\ref{A-Phi-a}) or~(\ref{tree-id}) between
$A_\mu$ and $\Phi$.

A general one-loop partial amplitude is obtained by gluing with a
scalar propagator the two extremities of the QCD dressed vertex
\begin{equation}
  \label{vertex-dressed}
  (- \mathrm{i} k_1^2) (- \mathrm{i} k_2^2) \Bigl\langle \chi(k_1)
  \chi(k_2) \Bigr\rangle_{1^+ \cdots n^+} =
  \raisebox{-14mm}{
  \begin{picture}(38,28)(-20,-14)
    \put(-20,-1){$\chi$}
    \put(-14, -2){\makebox(0,0){$\scriptstyle 1$}}
    \put(-14,0){\line(-1,0){2}}
    \put(0,0){\vector(-1,0){14}}
    \put(0,0){\vector(1,0){14}}
    \put(14,0){\line(1,0){2}}
    \put(14, 2){\makebox(0,0){$\scriptstyle 2$}}
    \put(18,-1){$\chi$}
    \put(-10,9){\makebox(0,0){tree}}
    \put(-10,0){\line(0,1){4.7}}\put(-10,9){\circle{8}}
    \put(-5,-9){\makebox(0,0){tree}}
    \put(-5,0){\line(0,-1){4.7}}\put(-5,-9){\circle{8}}
    \put(0,9){\makebox(0,0){tree}}
    \put(0,0){\line(0,1){4.7}}\put(0,9){\circle{8}}
    \put(10,-9){\makebox(0,0){tree}}
    \put(10,0){\line(0,-1){4.7}}\put(10,-9){\circle{8}}
  \end{picture}
  } \qquad ,
\end{equation}
or the SDYM dressed vertex $(- \mathrm{i} k_1^2) (- \mathrm{i} k_2^2)
\Bigl\langle \phi(k_1) \phi(k_2) \Bigr\rangle_{1 \cdots n}$. We found
in~(\ref{tree-id}) that there are no current amplitudes $ \Bigl\langle
A_{x+\mathrm{i}y} \Bigr\rangle$, so the two quartic vertices and the
third cubic vertex in~(\ref{vertex-QCD}) do not appear. For each
segment carrying one tree in the above diagram, we have (remember that
a current amplitude has a propagator attached to its off-shell leg):
\begin{eqnarray}
  \lefteqn{
  \raisebox{-3mm}{
  \begin{picture}(30,17)(-15,-3)
    \put(-15,-1){$\chi$}
    \put(-7, -2){\makebox(0,0){$\scriptstyle 1$}}
    \put(-7,0){\line(-1,0){3}}
    \put(0,0){\vector(-1,0){7}}
    \put(0,0){\vector(1,0){7}}
    \put(7,0){\line(1,0){3}}
    \put(7, -2){\makebox(0,0){$\scriptstyle 2$}}
    \put(12,-1){$\chi$}
    \put(0,9){\makebox(0,0){tree}}
    \put(0,0){\line(0,1){4.7}}\put(0,9){\circle{8}}
  \end{picture}
  }} \qquad \nonumber\\
  &&= \Bigl\langle A_{0+z}(k_3) \Bigr\rangle_{1^+ \cdots n^+}
  \raisebox{-8mm}{
  \begin{picture}(30,18)(-19,-9)
    \put(-5,0){\line(-1,0){5}}\put(0,0){\vector(-1,0){5}}
    \put(0,0){\vector(1,1){4}}\put(2.82,2.82){\line(1,1){4}}
    \put(0,0){\vector(1,-1){4}}\put(2.82,-2.82){\line(1,-1){4}}
    \put(-19,-1){$A_{0+z}$}\put(8,6){$\chi$}\put(8,-7.5){$\chi$}
    \put(1,-5){$\scriptstyle 1$}
    \put(.5,3){$\scriptstyle 2$}
    \put(-5,1){$\scriptstyle 3$}
  \end{picture}
  } \nonumber\\
  &&\quad + \Bigl\langle A_{x-\mathrm{i}y}(k_3) \Bigr\rangle_{1^+ \cdots n^+}
  \raisebox{-8mm}{
  \begin{picture}(31,18)(-20,-9)
    \put(-5,0){\line(-1,0){5}}\put(0,0){\vector(-1,0){5}}
    \put(0,0){\vector(1,1){4}}\put(2.82,2.82){\line(1,1){4}}
    \put(0,0){\vector(1,-1){4}}\put(2.82,-2.82){\line(1,-1){4}}
    \put(-20,-1){$A_{x-\mathrm{i}y}$}
    \put(8,6){$\chi$}\put(8,-7.5){$\chi$}
    \put(1,-5){$\scriptstyle 1$}
    \put(.5,3){$\scriptstyle 2$}
    \put(-5,1){$\scriptstyle 3$}
  \end{picture}
  }
  \nonumber\\
  &&= - \mathrm{i} \sqrt{2} \Bigl\langle \phi(k_3) \Bigr\rangle_{1 \cdots n}
  k_{3,x+\mathrm{i}y}
  \raisebox{-8mm}{
  \begin{picture}(30,18)(-19,-9)
    \put(-5,0){\line(-1,0){5}}\put(0,0){\vector(-1,0){5}}
    \put(0,0){\vector(1,1){4}}\put(2.82,2.82){\line(1,1){4}}
    \put(0,0){\vector(1,-1){4}}\put(2.82,-2.82){\line(1,-1){4}}
    \put(-19,-1){$A_{0+z}$}\put(8,6){$\chi$}\put(8,-7.5){$\chi$}
    \put(1,-5){$\scriptstyle 1$}
    \put(.5,3){$\scriptstyle 2$}
    \put(-5,1){$\scriptstyle 3$}
  \end{picture}
  } \nonumber\\
  &&\quad - \mathrm{i} \sqrt{2} \Bigl\langle \phi(k_3) \Bigr\rangle_{1 \cdots
    n} k_{3,0-z}
  \raisebox{-8mm}{
  \begin{picture}(31,18)(-20,-9)
    \put(-5,0){\line(-1,0){5}}\put(0,0){\vector(-1,0){5}}
    \put(0,0){\vector(1,1){4}}\put(2.82,2.82){\line(1,1){4}}
    \put(0,0){\vector(1,-1){4}}\put(2.82,-2.82){\line(1,-1){4}}
    \put(-20,-1){$A_{x-\mathrm{i}y}$}\put(8,6){$\chi$}
    \put(8,-7.5){$\chi$}
    \put(1,-5){$\scriptstyle 1$}
    \put(.5,3){$\scriptstyle 2$}
    \put(-5,1){$\scriptstyle 3$}
  \end{picture}
  }
  \nonumber\\
  &&= \Bigl\langle \phi(k_3) \Bigr\rangle_{1 \cdots n}
  \raisebox{-8mm}{
  \begin{picture}(25,18)(-14,-9)
    \put(-5,0){\line(-1,0){5}}\put(0,0){\vector(-1,0){5}}
    \put(0,0){\vector(1,1){4}}\put(2.82,2.82){\line(1,1){4}}
    \put(0,0){\vector(1,-1){4}}\put(2.82,-2.82){\line(1,-1){4}}
    \put(-14,-1){$\phi$}\put(8,6){$\phi$}\put(8,-7.5){$\phi$}
    \put(1,-5){$\scriptstyle 1$}
    \put(.5,3){$\scriptstyle 2$}
    \put(-5,1){$\scriptstyle 3$}
  \end{picture}
  } =
  \raisebox{-3mm}{
  \begin{picture}(30,17)(-15,-3)
    \put(-15,-1){$\phi$}
    \put(-7, -2){\makebox(0,0){$\scriptstyle 1$}}
    \put(-7,0){\line(-1,0){3}}
    \put(0,0){\vector(-1,0){7}}
    \put(0,0){\vector(1,0){7}}
    \put(7,0){\line(1,0){3}}
    \put(7, -2){\makebox(0,0){$\scriptstyle 2$}}
    \put(12,-1){$\phi$}
    \put(0,9){\makebox(0,0){tree}}
    \put(0,0){\line(0,1){4.7}}\put(0,9){\circle{8}}
  \end{picture}
  } \,.
\end{eqnarray}
For the second equality, we use the tree level
identity~(\ref{tree-id}) relating $\Bigl\langle A(k) \Bigr\rangle_{1^+
  \cdots n^+}$ and $\Bigl\langle \phi(k) \Bigr\rangle_{1 \cdots n}$.
For the third equality, we replace a linear combination of two QCD
vertices~(\ref{vertex-QCD}) with a $\phi$-vertex~(\ref{vertex-phi}).
The dressed vertices of scalar QCD and of the
SDYM action coincide and their one-loop amplitudes are therefore
equal.

These amplitudes have been explicitly calculated in the context of QCD
for an arbitrary number of legs: their form has first been
conjectured~\cite{gordon} using gluons in the loop and then
derived~\cite{mahlon-loop} using fermions in the loop. The result is
amazingly simple:
\begin{eqnarray}
  \lefteqn{
  (- \mathrm{i} k_1^2) \cdots (- \mathrm{i} k_n^2)
  \left. \Bigl\langle \phi(k_1) \cdots \phi(k_n) \phi(k) \Bigr\rangle
  \right|_{k_1^2 = \cdots = k_n^2 = 0} \; \rlap{\ \ \ \
    $=$}\raisebox{1em}{one-loop} \; - \frac{N_c \, g^n }{2}\,
  \frac{\mathrm{i}^{n+1}}{48 \pi^2} } \quad \nonumber\\
  &&\times
  \sum_{i=2}^{n-2} \sum_{j=i+1}^{n-1} \frac{k_{i,\mu} \, (k_1 + \cdots
    + k_i)_\nu \, k_{j,\lambda} \, (k_1 + \cdots + k_j)_\rho}{(Q_1 -
    Q_2) (Q_2 - Q_3) \cdots (Q_n - Q_1)} \; \;\mathrm{tr}\;
  \bar\sigma^\mu \sigma^\nu \bar\sigma^\lambda \sigma^\kappa \,,
\end{eqnarray}
with the Pauli matrices, $\sigma^\mu = (1, \sigma^i), \bar\sigma^\mu
= (1, - \sigma^i)$.

I end this section with some comments on the alternative SDYM
action~(\ref{actionJ}). Since its equations of motion are also the
SDYM ones, this action reproduces correctly the tree amplitudes.
At one-loop, the proliferation of vertices makes the analysis more
difficult. I checked that the four particle amplitude gives
again~(\ref{QCD-1L}). With $J = \exp \mathrm{i} \pi^a T^a / f_\pi$,
the cubic vertex is a slight modification of ~(\ref{vertex-phi}):
\begin{equation}
  \label{vertex-pi}
  \raisebox{-8mm}{
  \begin{picture}(25,18)(-14,-9)
    \put(-5,0){\line(-1,0){5}}\put(0,0){\vector(-1,0){5}}
    \put(0,0){\vector(1,1){4}}\put(2.82,2.82){\line(1,1){4}}
    \put(0,0){\vector(1,-1){4}}\put(2.82,-2.82){\line(1,-1){4}}
    \put(-14,-1){$\pi$}\put(8,6){$\pi$}\put(8,-7.5){$\pi$}
    \put(1,-5){$\scriptstyle 1$}
    \put(.5,3){$\scriptstyle 2$}
    \put(-5,1){$\scriptstyle 3$}
  \end{picture}
  } = - \frac{1}{6 f_\pi} \; \biggl( X(k_3, k_1 - k_2) + k_1^2 - k_2^2
    \biggr) \,.
\end{equation}
The additional term $( k_1^2 - k_2^2 )$ is irrelevant for the
four particle amplitude, but does contribute to higher amplitudes, as
do the quartic, quintic, \textit{etc.} vertices. This also happens at
tree level but does not change the final result. It is plausible that
the same cancellations work at one-loop and that the
Donaldson--Nair--Schiff action reproduces also all the one-loop MHV
amplitudes.

\section{Symmetries of the SDYM system}

In fact, the dressed vertices~(\ref{vertex-dressed}) have an
interesting interpretation in terms of the symmetries of the SDYM
equations (see also~\cite{selivanov}). Consider a small perturbation
$\Lambda(x)$ around a solution $\Phi$ of the SDYM equations. It
satisfies the linearization of Eq.~(\ref{equPhi}),
\begin{equation}
  \label{equLambda}
  \partial^2 \Lambda - \mathrm{i} g \; [ \partial_{x+\mathrm{i}y} \Lambda,
  \partial_{0-z} \Phi ] - \mathrm{i} g \; [ \partial_{x+\mathrm{i}y} \Phi,
  \partial_{0-z} \Lambda ] = 0 \,.
\end{equation}
A similar equation is obtained by differentiating~(\ref{equPhiJ}) with
respect to $J(y)$:
\begin{eqnarray}
  \label{equLambdaJ}
  \lefteqn{
  \partial^2 \frac{\delta\Phi_J(x)}{\delta J(y)} - \mathrm{i} g \; [
  \partial_{x+\mathrm{i}y} \frac{\delta\Phi_J(x)}{\delta J(y)},
  \partial_{0-z} \Phi_J ] - \mathrm{i} g \; [ \partial_{x+\mathrm{i}y}
  \Phi_J, \partial_{0-z} \frac{\delta\Phi_J(x)}{\delta J(y)} ] }
  \qquad \hspace{3cm} \nonumber\\
  &&{} + \delta(x-y) = 0 \,.
\end{eqnarray}
Actually, $\Lambda_q(k) \equiv (- i q^2) \; \lambda \; \delta
\Phi_J(k) / i\delta J(q)$ is nothing else than the generating
functional of the dressed vertex $(- \mathrm{i} k^2) (- \mathrm{i}
q^2) \lambda \Bigl\langle \phi(k) \phi(q) \Bigr\rangle_{1 \cdots n}$.
The constant $\lambda$ will be fixed later. We can solve for
$\Lambda_q(k)$ as we did in section~3.  Take $\Phi$ to be the Bethe
\textit{Ansatz} solution but with the plane wave $j=j*$ missing ($j*$
is some index between 1 and $n$) and expand $\Lambda_{q=k_{j*}}$ in
powers of the coupling constant $g$:
\begin{eqnarray}
  \displaystyle
  &\displaystyle \Phi(x) = \sum_{m=1}^\infty \Phi^{(m)}(x) \qquad
  \mathrm{with} \qquad \Phi^{(1)}(x) = - \mathrm{i} \sum_{j=1\atop j
    \not= j*}^n T^{a_j} \e^{-\mathrm{i} k_j x} f(k_j) \,,
  \nonumber\\[-6pt]
  \\[-6pt]
  &\displaystyle \Lambda_{j*}(x) = \sum_{m=1}^\infty \Lambda^{(m)}_{j*}(x)
  \qquad \mathrm{with} \qquad \Lambda^{(1)}_{j*}(x) = - \mathrm{i}
  T^{a_{j*}} \e^{-\mathrm{i} k_{j*} x} \,. \nonumber
\end{eqnarray}
The function $f(k_j)$ has support on the light-cone $k_j^2 = 0$.
However $k_{j*}$ is unrestricted and may be off-shell.\footnote{A
  solution of the homogeneous equation~(\ref{equLambda}) is obtained
  by restricting $k_{j*}$ to be on-shell.} At successive orders in the
coupling constant $g$, we have diagrammatically
\begin{eqnarray}
  \label{linear-rec}
  \Lambda^{(m)}_{j*}(k) = \mathrm{i}g\sum_{j =1}^{j*-1}
  \raisebox{-8mm}{
  \begin{picture}(37,18)(-10,-9)
    \put(-5,0){\line(-1,0){5}}\put(0,0){\vector(-1,0){5}}
    \put(0,0){\vector(1,1){4}}\put(2.82,2.82){\line(1,1){4}}
    \put(0,0){\vector(1,-1){4}}\put(2.82,-2.82){\line(1,-1){4}}
    \put(8,6){$\Lambda^{(m-j)}_{j*}(k^{\prime\prime})$}
    \put(8,-7.5){$\Phi^{(j)}(k')$}
  \end{picture}
  } + \mathrm{i}g\sum_{j=j*}^{m-1}
  \raisebox{-8mm}{
  \begin{picture}(37,18)(-10,-9)
    \put(-5,0){\line(-1,0){5}}\put(0,0){\vector(-1,0){5}}
    \put(0,0){\vector(1,1){4}}\put(2.82,2.82){\line(1,1){4}}
    \put(0,0){\vector(1,-1){4}}\put(2.82,-2.82){\line(1,-1){4}}
    \put(8,6){$\Phi^{(m-j)}(k^{\prime\prime})$}
    \put(8,-7.5){$\Lambda^{(j)}_{j*}(k')$}
  \end{picture}
  } \,, \nonumber\\
\end{eqnarray}
The trees in $\Lambda^{(n)}_{j*}(k)$ have $(n-1)$ on-shell legs and two
off-shell legs with momenta $k$ and $k_{j*}$. The sum of the trees
ending with $n$ different momenta $k_j$ ($j = 1, \ldots, n$)
reproduces the dressed vertex $( - \mathrm{i} k_{j*}^2) \lambda
\Bigl\langle \phi(k) \phi(k_{j*}) \Bigr\rangle_{1 \cdots
  \rlap{$\!\backslash$} j* \cdots n}$ with two off-shell momenta $k$
and $k_{j*}$. One can even find a closed analytic expression for it;
for $f(k) = Q / k_{x+\mathrm{i}y}$ as before and $\lambda =
- 2$, equations~(\ref{linear-rec}) are precisely the equations that
Dunn, Mahlon and Yan have for two off-shell current amplitudes in
QCD~\cite{mahlon}.  Explicit expressions can be found in their article
and will not be reproduced here.

A one-parameter family of symmetries $\Lambda_s$ is constructed from
the following pair of recursion relations~\cite{dolan}:
\begin{eqnarray}
  &\partial_{0-z} \Lambda_{s+1} = \partial_{x-\mathrm{i}y} \Lambda_s -
  \mathrm{i} g [ \partial_{0-z} \Phi, \Lambda_s ] \,,
  \nonumber\\[-6pt]
  \label{hierarchy} \\[-6pt]
  &\partial_{x+\mathrm{i}y} \Lambda_{s+1} = \partial_{0+z} \Lambda_s -
  \mathrm{i} g [ \partial_{x+\mathrm{i}y} \Phi, \Lambda_s ]
  \,. \nonumber
\end{eqnarray}
These equations are compatible if $\Lambda_s$ is a solution
of~(\ref{equLambda}). Moreover, $\Lambda_{s+1}$ is a symmetry if
$\Phi$ is a solution of the SDYM equations. It is known that these
symmetries form an affine Lie algebra. This is at the basis of the
conjecture~\cite{green} that all integrable models are some reduction
of a SDYM system.  Of course, the $\Lambda$'s are symmetries of the
equations of motion and not necessarily of the action. In fact, in the
hierarchy~(\ref{hierarchy}), only $\Lambda_0 = T^a$ and $\Lambda_1 =
-\mathrm{i} g \; [ \phi, T^a ]$ are true symmetries of the action.
Nevertheless, the hierarchy~(\ref{hierarchy}) defines an infinite set
of conserved currents whose classical expressions are
\begin{eqnarray}
  &\mathcal{J}_{s,0+z} = \partial_{0+z} \Lambda_s - \mathrm{i}g\; [
  \partial_{x+\mathrm{i}y} \Phi, \Lambda_s ] \,, \quad
  \mathcal{J}_{s,x-\mathrm{i}y} = \partial_{x-\mathrm{i}y} \Lambda_s -
  \mathrm{i}g\; [ \partial_{0-z} \Phi, \Lambda_s ] \,, \nonumber\\
  &\partial_\mu \mathcal{J}^\mu_s = \frac{1}{2} \mathrm{i}g\; [ \Lambda_{s-1},
  \partial^2 \Phi - \mathrm{i}g\; [ \partial_{x+\mathrm{i}y} \Phi,
  \partial_{0-z} \Phi ] = 0 \,.
\end{eqnarray}
Since these currents are only known for classical solutions $\Phi$,
one can only derive tree level identities and not true Ward identities
relating Green functions of different orders in $\hbar$.

\section{Conclusion}

In this paper, I have investigated the idea that QCD computations may
be simplified when considering only maximally helicity (MHV) violating
processes. Although this statement seems natural, its concrete
realization is based upon special tools like supersymmetry identities
and self-dual Yang--Mills (SDYM) equations. I showed that a scalar
action reproduces exactly the MHV amplitudes up to one loop. The
extension of this result beyond one-loop is not obvious, since I can
no more invoke supersymmetric arguments to restrict myself to scalar
QCD. It is unlikely that the SDYM action is sufficient to describe the
MHV amplitudes at two-loop, even when all external helicities are set
equal. Namely, any cut in a diagram that isolates a tree amplitude
will be automatically zero for the SDYM action, whereas it will be
non-zero in QCD owing to the presence of different helicities in the
intermediate states.

The vanishing of tree amplitudes is the signature of a quantized
integrable model~\cite{ooguri}. We have further probed Bardeen's idea
that the $S$-matrix of QCD between positive helicity gluons is
intimately related to the quantization of a SDYM system. The fact that
one-loop amplitudes are calculated from a generator of the symmetry of
the SDYM equations is one more piece of evidence in favour of an
anomaly-type mechanism to generate the simple non-vanishing one-loop
amplitudes. A more detailed investigation of this question is still
needed.

A necessary step is to study the quantum symmetries of the system. It
was noted at the end of the previous section that the SDYM action has
global symmetries and that classically an infinite set of conserved
currents can be found. It is still an open question whether they are
responsible for the vanishing of tree amplitudes.  The
Donaldson--Nair--Schiff action~(\ref{actionJ}) may be an interesting
alternative, since it also possesses an infinite-dimensional current
algebra symmetry. This model is moreover viewed~\cite{moore} as an
example of a four-dimensional conformal theory and it was already
suggested in Ref.~\cite{ooguri} that the infinite number of symmetries
generates Ward identities that ensure the vanishing of tree
amplitudes. We have checked that the Donaldson--Nair--Schiff action
reproduces the correct QCD tree current amplitudes and one-loop
four particle amplitude.  However, due to a proliferation of vertices,
we have been unable so far to show that it also reproduces the other
one-loop amplitudes. If this proves to be true, then all the one-loop
amplitudes of this conformal theory may be
deduced\footnote{Incidentally, there is a delicate issue with the
  signature of spacetime. The Donaldson--Nair--Schiff action and the
  SDYM equations are usually written using a $(2+2)$ signature. One
  must be careful with analytic continuation; for example, I already
  mentioned that the cubic vertex within this signature does not
  vanish.} from available QCD calculations~\cite{gordon,mahlon-loop}.

There remains several other open questions.  What can this approach
bring in return to QCD? Can the mixing of helicities be described by
some appropriate coupling between a self-dual and an anti-self-dual
fields?  These questions deserve further investigation.

\begin{ack}
  I thank Zvi Bern for pointing out Ref.~\cite{bardeen} to me and for
  stimulating discussions, which motivated the present work. The
  program I used to compute the four-point function was written by
  Andrew Morgan. I thank him and Lance Dixon for many useful comments.
\end{ack}

\section*{Note added}

After completion of this work, two related papers appeared. Chalmers
and Siegel~\cite{siegel} propose yet another SDYM action $S = \int
d^4x \; \Lambda \, ( \partial^2 \phi - \mathrm{i} g [
\partial_{x+\mathrm{i}y} \phi, \partial_{0-z} \phi ] )$. It is derived
as a truncation of the $N=4$ supersymmetric Yang--Mills action in the
light-cone formalism.  The equations of motion for the Lagrange
multiplier $\Lambda$ are identical to the equations~(\ref{equLambda})
for the generators of the SDYM symmetries. Moreover, the authors prove
my conjecture that the Donaldson-Nair-Schiff action gives the correct
one-loop amplitudes. Korepin and Oota~\cite{oota} derive a closed
expression for the Bethe Ansatz solution described in Appendix~B.

\appendix

\section{Spinor notations}

This appendix is intended for readers familiar with the spinor
notation. Consider first a massless Weyl spinor
\begin{equation}
  u^\alpha(k) = \left( \frac{ k_{x+\mathrm{i}y} }{ \sqrt{|k_{0+z}|} } , -
  \mathrm{sign} (k_0) \sqrt{|k_{0+z}|} \right) \,.
\end{equation}
Notice that on-shell we have either $k_{0+z}, k_{0-z}, k_0 > 0$ or
$k_{0+z}, k_{0-z}, k_0 < 0$. The spinor products
\begin{eqnarray}
  &\langle 1 2 \rangle = u^\alpha(k_1) u_\alpha(k_2) = - (Q_1 - Q_2)
   \frac{k_{1,x+\mathrm{i}y}}{\sqrt{|k_{1,0+z}|}}
   \frac{k_{2,x+\mathrm{i}y}}{\sqrt{|k_{2,0+z}|}}
  \nonumber\\[-6pt]
  \label{spinor} \\[-6pt]
  &[12] = \mathrm{sgn} \; ( k_1^0 k_2^0 ) \langle 2 1 \rangle^* =
  X(k_1,k_2)
  \frac{\sqrt{|k_{1,0+z}|}}{k_{1,x+\mathrm{i}y}}
  \frac{\sqrt{|k_{2,0+z}|}}{k_{2,x+\mathrm{i}y}}
  \,, \nonumber\\
\end{eqnarray}
satisfy
\begin{equation}
  2\; k_1 \cdot k_2 = \langle 1 2 \rangle \, [ 2 1 ] \,.
\end{equation}

One defines polarization vectors in a reference frame $q$ by their
projection on an arbitrary null vector $p$,
\begin{equation}
  p \cdot \varepsilon^+(k;q) = \frac{\langle q p
    \rangle [p k]}{\sqrt{2} \; \langle q k \rangle} \; \mathrm{sign}
    (k_0) \,, \quad
  p \cdot \varepsilon^-(k;q) = \frac{\langle k p
    \rangle [p q]}{\sqrt{2} \; [k q]} \; \mathrm{sign} (k_0) \,.
\end{equation}
(Notice the opposite sign in Ref.~\cite{mahlon}.)

The four point function in these notations takes the compact form
\begin{equation}
  m^{\mathrm{one-loop}}(k_1,+; \cdots k_4,+) = - N_c \, g^4
  \frac{\mathrm{i}}{48 \pi^2} \frac{[ 12 ] [ 34 ]}{\langle 12 \rangle
  \langle 34 \rangle} \,,
\end{equation}
which is the one usually quoted in the literature.

\section{The Bethe \textit{Ansatz} for a SDYM solution}

In this appendix, I indicate why~(\ref{treePhiJ}) is indeed a solution
and I comment on the terms left over in the Bethe \textit{Ansatz}
solution~(\ref{treePhi}).

It is easy to see that the recursion relation~(\ref{recrel}) is
satisfied by~(\ref{treePhiJ}) provided the following identity is valid
on-shell:
\begin{eqnarray}
  (p_1 + \cdots + p_m)^2 = \sum_{j=1}^{m-1} X(p_1 + \cdots +
  p_j, p_{j+1} + \cdots + p_m) (Q_j - Q_{j+1}) \,.
\end{eqnarray}
The proof goes as follow:
\begin{eqnarray}
  \lefteqn{
    \sum_{j=1}^{m-1} X(p_1 + \cdots + p_j, p_{j+1} + \cdots  p_m) \; (Q_j
    - Q_{j+1}) } \qquad \nonumber\\
  && = \sum_{j=1}^{m-1} \sum_{s=1}^j \sum_{t=j+1}^m X(p_s, p_t) \; (Q_j -
    Q_{j+1}) \nonumber\\
  && = \sum_{1 \leq s \leq j < t \leq m} 2 (p_s \cdot p_t) \;
    (Q_s - Q_t)^{-1} (Q_j - Q_{j+1}) \nonumber\\
  && = \sum_{s=1}^m \sum_{t=s+1}^m 2 (p_s \cdot p_t) \; (Q_s - Q_t)^{-1}
    \; \underbrace{\sum_{j=s}^{t-1} \; (Q_j - Q_{j+1})}_{=(Q_s - Q_t)}
    \nonumber\\
  && = \sum_{s=1}^m \sum_{t=s+1}^m 2 (p_s \cdot p_t) = (p_1 + \cdots +
    p_m)^2 \,,
\end{eqnarray}
and is based on the algebraic relations~(\ref{Xdef}) and~(\ref{Xprop}).

After inserting the plane wave expansion~(\ref{planew}) for $j(x)$
into~(\ref{treePhiJ}), the $p_j$'s momenta are replaced by the plane
wave momenta. It may happen that two or more consecutive $p_j,
p_{j+1}$ are replaced by the same momentum $k_\ell$. In that case one may
worry about factors like $(Q_j - Q_{j+1})^{-1}$. The way to deal with
this problem is first to replace $Q_j$ by $Q_\ell + \epsilon_j$, then to
symmetrize all the $\epsilon_j$'s and finally to take the limit
$\epsilon_j \to 0$.  For example, the limits
\begin{eqnarray*}
  (Q_1 - Q_2)^{-1} (Q_2 - Q_3)^{-1} \quad
     \raisebox{8pt}{$\scriptscriptstyle p_1=p_2\not=p_3$}
     \hspace{-11mm}&\longrightarrow&\hspace{4mm}
     \frac{1}{2} (Q_1 - Q_3)^{-2} \,, \\
  (Q_1 - Q_2)^{-1} (Q_2 - Q_3)^{-1} \quad
     \raisebox{8pt}{$\scriptscriptstyle p_1=p_2=p_3$}
     \hspace{-11mm}&\longrightarrow&\hspace{4mm}
     0 \,, \\
  (Q_1 - Q_2)^{-1} (Q_2 - Q_3)^{-1} (Q_3 - Q_4)^{-1} \quad
     \raisebox{8pt}{$\scriptscriptstyle p_1\not=p_4 \,,\, p_2=p_3$}
     \hspace{-13mm}&\longrightarrow&\hspace{6mm}
     \frac{1}{2} (Q_1 - Q_4) (Q_1 - Q_2)^{-2} \nonumber\\
     &&{} \qquad \times (Q_2 - Q_4)^{-2} \,, \\
  (Q_1 - Q_2)^{-1} (Q_2 - Q_3)^{-1} (Q_3 - Q_4)^{-1} \quad
     \raisebox{8pt}{$\scriptscriptstyle p_1=p_2\not=p_3=p_4$}
     \hspace{-13mm}&\longrightarrow&\hspace{6mm}
     \frac{1}{2} (Q_1 - Q_4)^{-3} \,, \\
  (Q_1 - Q_2)^{-1} (Q_2 - Q_3)^{-1} (Q_3 - Q_4)^{-1} \quad
     \raisebox{8pt}{$\scriptscriptstyle p_1=p_4 \,,\, p_2=p_3$}
     \hspace{-13mm}&\longrightarrow&\hspace{7mm}
     0 \,, \\
  (Q_1 - Q_2)^{-1} (Q_2 - Q_3)^{-1} (Q_3 - Q_4)^{-1} \quad
     \raisebox{8pt}{$\scriptscriptstyle p_1=p_2=p_3\not=p_4$}
     \hspace{-13mm}&\longrightarrow&\hspace{7mm}
     \frac{1}{6} (Q_1 - Q_4)^{-3} \,, \\
\end{eqnarray*}
enter in the expressions for the first few terms, $\Phi^{(1)}$,
$\Phi^{(2)}$, $\Phi^{(3)}$ and $\Phi^{(4)}$, in the power expansion of
the Bethe \textit{Ansatz} solution:
\begin{eqnarray}
  && \Phi^{(1)}(x) = - \mathrm{i} \sum_{j=1}^n \; T^{a_j}
  \e^{-\mathrm{i} k_j x} f(k_j) \,, \\
  && \Phi^{(2)}(x) = - \mathrm{i} g \sum_{ \mathrm{permutations\ of\ } \atop
     ( 1 2 ) \mathrm{\ in\ } \{ 1 \cdots n \} } T^{a_1} T^{a_2} \e^{ -
     \mathrm{i} (k_1 + k_2) x } f(k_1) f(k_2) (Q_1 - Q_2)^{-1} \,, \\
  && \Phi^{(3)}(x) = - \mathrm{i} g^2 \sum_{ \mathrm{permutations\ of\ } \atop
     ( 1 2 3 ) \mathrm{\ in\ } \{ 1 \cdots n \} } T^{a_1} T^{a_2} T^{a_3}
     \e^{ - \mathrm{i} (k_1 + k_2 + k_3) x } f(k_1) f(k_2) f(k_3)
     \nonumber\\
     && \hspace{6cm} {} \times (Q_1 - Q_2)^{-1} (Q_2 - Q_3)^{-1} \nonumber\\
  &&{} \quad - \mathrm{i} \frac{g^2}{2} \sum_{ \mathrm{permutations\
  of\ } \atop ( 1 2 ) \mathrm{\ in\ } \{ 1 \cdots n \} } T^{a_1}
  T^{a_1} T^{a_2} \e^{ - \mathrm{i} (2 k_1 + k_2) x } f(k_1)^2 f(k_2)
  (Q_1 - Q_2)^{-2} \nonumber\\
  &&{} \quad + \mathrm{i}  g^2 \sum_{ \mathrm{permutations\ of\ } \atop
     ( 1 2 ) \mathrm{\ in\ } \{ 1 \cdots n \} } T^{a_1} T^{a_2}
     T^{a_1} \e^{ - \mathrm{i} (2 k_1 + k_2) x } f(k_1)^2 f(k_2) (Q_1
     - Q_2)^{-2} \nonumber\\
  &&{} \quad - \mathrm{i} \frac{g^2}{2} \sum_{ \mathrm{permutations\
  of\ } \atop ( 1 2 ) \mathrm{\ in\ } \{ 1 \cdots n \} } T^{a_1}
  T^{a_2} T^{a_2} \e^{ - \mathrm{i} (k_1 + 2 k_2) x } f(k_1) f(k_2)^2
  (Q_1 - Q_2)^{-2}
\end{eqnarray}
It would be interesting to get a closed formula for all $\Phi^{(m)}(x)$.


\begin{thebibliography}{99}

\bibitem{bern} For a recent review and a list of references see:
  Z.~Bern, L.~Dixon and D.A.~Kosower, \textit{Progress in One-Loop QCD
    Computations}, to appear in Ann. Rev. Nucl. Part. Sci. (1996).

\bibitem{parke}
  S.J.~Parke and T.~Taylor, Nucl. Phys. \textbf{B269} (1986) 410.

\bibitem{berends}
  F.A.~Berends and W.T.~Giele, Nucl. Phys. \textbf{B306} (1988) 759.

\bibitem{mangano}
  M.~Mangano and S.J.~Parke, Phys. Rep. \textbf{200} (1991) 301.

\bibitem{ooguri}
  H.~Ooguri and C.~Vafa, Nucl. Phys. \textbf{B367} (1991) 83.

\bibitem{korepin} I.Y.~Aref'eva and V.E.~Korepin, JETP
  Lett. \textbf{20} (1974) 312.

\bibitem{nair}
  V.P.~Nair, Phys. Lett. B \textbf{214} (1988) 215.

\bibitem{bardeen}
  W.A.~Bardeen, \textit{Self-dual Yang--Mills Theory, Integrability and
    Multi-Parton Amplitudes}, FERMILAB-CONF-95-379-T, presented at
  Yukawa International Seminar '95: From the Standard Model to Grand
  Unified Theories, Kyoto, Japan, 21-25 Aug 1995.

\bibitem{selivanov}
  K.G.~Selivanov, \textit{Multigluon Tree Amplitude and Self-Duality
    Equation}, ITEP-21/96, hep-ph/9604206.

\bibitem{spinor}
  F.A.~Berends, R.~Kleiss, P.~De~Causmaecker, R.~Gastmans and T.T.~Wu,
  Phys. Lett. B \textbf{103} (1981) 124;
  P.~De~Causmaecker, R.~Gastmans, W.~Troost and T.T.~Wu, Nucl. Phys \textbf{
    B206} (1982) 53;
  R.~Kleiss and W.J.~Stirling, Nucl. Phys. \textbf{B262} (1985) 235;
  J.F.~Gunion and Z.~Kunszt, Phys. Lett. B \textbf{161} (1985) 333;
  Z.~Xu, D.-H.~Zhang and L.~Chang, Nucl. Phys. \textbf{B291} (1987) 392.

\bibitem{mahlon} G.D.~Mahlon, T.M.~Yan and C.~Dunn, Phys. Rev. D \textbf{48}
  (1993) 1337.

\bibitem{tasi}
  Z.~Bern, in \textit{Proceedings of Theoretical Advanced Study Institute
    in High Energy Physics (TASI 92)}, eds. J.~Harvey and
  J.~Polchinski (World Scientific, Singapore, 1993).

\bibitem{andrew}
  Z.~Bern and A.G.~Morgan, Phys. Rev. D \textbf{49} (1994) 6155.

\bibitem{color}
  J.E.~Paton and Chan~Hong-Mo, Nucl. Phys. \textbf{B10} (1969) 519;
  F.A.~Berends and W.T~Giele, Nucl. Phys. \textbf{B294} (1987) 700;
  M.~Mangano, S.~Parke and Z.~Xu, Nucl. Phys. \textbf{B298} (1988) 653;
  M.~Mangano, Nucl. Phys. \textbf{B309} (1988) 461;
  Z.~Bern and D.A.~Kosower, Nucl. Phys. \textbf{B362} (1991) 389.

\bibitem{dixon}
  Z.~Bern, L.~Dixon, D.C.~Dunbar and D.A.~Kosower, Nucl. Phys. \textbf{
    B425} (1994) 217;
  Z.~Bern, L.~Dixon and D.A.~Kosower, Nucl. Phys. \textbf{B437} (1995)
  259.

\bibitem{david}
  Z.~Bern and D.A.~Kosower, Nucl. Phys. \textbf{B379} (1992) 451.

\bibitem{yang} C.N.~Yang, Phys. Rev. Lett. \textbf{38} (1977) 1377.

\bibitem{donaldson} S.~Donaldson, Proc. Lond. Math. Soc. \textbf{50}
  (1985) 1.

\bibitem{schiff} V.P.~Nair and J.~Schiff, Phys. Lett. B \textbf{246} (1990)
  423.

\bibitem{moore} A.~Losev, G.~Moore, N.~Nekrasov and S.~Shatashvili,
  \textit{Four-Dimensional Avatars of Two-Dimensional RCFT}, PUPT-1564,
  talk given at Strings 95, Los Angeles, CA, March 13-18, 1995 and
  Trieste Conference on S Duality and Mirror Symmetry, Trieste, Italy, June
  5-6, 1995, hep-th/9509151.

\bibitem{ketov} S.~Ketov, \textit{All Loop Finiteness of the
    Four-Dimensional Donaldson--Nair--Schiff Non-Linear Sigma-Model},
  DESY 96-071 and ITP-UH-05/96, hep-th/9604141.

\bibitem{bruschi} M.~Bruschi, D.~Levi and O.~Ragnisco,
  Lett. Nuov. Cim. \textbf{33} (1982) 263.

\bibitem{parkes} A.N~Leznov and M.A.~Mukhtarov, J. Math. Phys. \textbf{28}
  (1987) 2574; A.~Parkes, Phys. Lett. B \textbf{286} (1992) 265.

\bibitem{recursion}
  D.A.~Kosower, Nucl. Phys. \textbf{B335} (1990) 23;
  G.D.~Mahlon, Phys. Rev. D \textbf{49} (1994) 2197.

\bibitem{gordon}
  Z.~Bern, G.~Chalmers, L.~Dixon and D.A.~Kosower, Phys. Rev. Lett. \textbf{
    72} (1994) 2134;
  Z.~Bern, L.~Dixon and D.A.~Kosower, \textit{Proceedings of Strings 1993},
  eds. M.~Halpern, A.~Sevrin and G.~Rivlis (World Scientific,
  Singapore, 1994).

\bibitem{mahlon-loop}
  G.D.~Mahlon, Phys. Rev. D \textbf{49} (1994) 4438.

\bibitem{dolan}
  M.K~Prasad, A.~Sinha and L.-L.~Wang, Phys. Lett. \textbf{87B} (1979) 237;
  L.~Dolan, Phys. Rep. \textbf{109} (1984) 1.

\bibitem{green}
  R.S.~Ward, Phil. Trans. Roy. Soc. Lond. A \textbf{315} (1985) 451.

\bibitem{siegel}
  G.~Chalmers and W.~Siegel, \textit{The Self-Dual Sector of QCD
  Amplitudes}, preprint ITP-SB-96-29, June 1996.

\bibitem{oota}
  V.E.~Korepin and T.~Oota, \textit{Scattering of Plane Waves in
  Self-Dual Yang--Mills Theory}, preprint YITP-96-33, August 1996.
\end{thebibliography}
\end{document}